\documentclass[twocolumn]{aastex701}
\usepackage{amsmath}

\begin{document}

\title{Impact of disk magnetic fields on the propagation of stellar-scale jets in the magnetically arrested accretion disks of active galactic nuclei}

\author[orcid=0000-0002-4448-0849,sname=Huang,gname=Bao-Quan]{Bao-Quan Huang}
\affiliation{College of Intelligent Manufacturing, Nanning University, Nanning, Guangxi 530299, People's Republic of China}
\email{huangbaoquan@unn.edu.cn}

\author[orcid=0000-0001-8678-6291,sname=Liu,gname=Tong]{Tong Liu}\thanks{E-mail: tongliu@xmu.edu.cn}
\affiliation{Department of Astronomy, Xiamen University, Xiamen, Fujian 361005, People's Republic of China}
\email{tongliu@xmu.edu.cn}

\author[sname=Zhang,gname=Jian-Fu]{Jian-Fu Zhang}
\affiliation{Department of Physics, Xiangtan University, Xiangtan, Hunan 411105, People's Republic of China}
\email{jfzhang@xtu.edu.cn}

\begin{abstract}
It is widely recognized that active galactic nucleus (AGN) disks host numerous massive stars and compact objects. Stellar-scale jets triggered by collapses of massive stars and mergers of compact objects could propagate through the disk and produce observable electromagnetic radiation. Magnetically arrested disks (MADs), supported by both numerical simulations and observations, possess strong magnetic fields (MFs). As jets travel within such environments, the MFs should regulate jet evolution and shape radiation signatures. In this work, we explore the effects of disk MFs on jet propagation and breakout emission within the MAD framework. We employ a jet-cocoon model that accounts for potential disk-MF effects, including both magnetic pressure and magnetic energy dissipation driven by magnetic reconnection. We find that magnetic pressure effectively suppresses the lateral expansion of the cocoon, which enhances jet collimation and modestly increases the jet-head velocity. Furthermore, magnetic pressure effects are more pronounced at relatively low jet powers. In this regime, the breakout luminosity of the jet-head shock is enhanced, while its breakout time is shortened. However, the magnitude of the luminosity enhancement is sensitive to the adopted regime-dependent emission prescriptions. These findings suggest that, within the explored parameter space, disk MFs can facilitate the breakout of low-power jets arising from binary black hole mergers in AGN MADs.
\end{abstract}


\section{Introduction}
It is generally believed that accretion disks of active galactic nuclei (AGNs) harbor abundant massive stars and compact objects. These massive stars and compact objects are thought to originate either through capture from nuclear star clusters \citep[e.g.,][]{Artymowicz1993ApJ,Fabj2020MNRAS,Nasim2023MNRAS,Wang2024MNRAS} or through in situ formation within the disk \citep[e.g.,][]{Goodman2003MNRAS,Dittmann2020MNRAS,Fan2023ApJ,Fabj2025ApJ}. Collapses of massive stars and mergers of compact-object binaries can trigger gamma-ray bursts (GRBs). As GRB jets propagate through and eventually break out of the disk, they can generate observable electromagnetic emission. In general, the emission is expected to be dominated by thermal radiation, appearing as optical, UV, or X-ray flares, with durations ranging from several hours to tens of days, and luminosities spanning a broad range of  $\sim10^{40}$--$10^{46}\,\mathrm{erg/s}$ \citep[e.g.,][]{McKernan2019ApJ,Wang2021ApJ,Zhu2021ApJL,Tagawa2023ApJb,Tagawa2024ApJ,Chen2024ApJ,Xing2025ApJ,Chen2025ApJ,Ma2025PRD,Rodr2025PhRvD,Tagawa2026arXiv}. However, if jet velocities remain relativistic after successfully breaking out of the disk, nonthermal emission is also expected \citep[e.g.,][]{Perna2021ApJL,Yuan2022ApJ,Lazzati2022ApJL,Lazzati2023ApJL,Wang2022MNRAS,Ray2023MNRAS,Kathirgamaraju2024ApJ,Huang2024ApJ,Yuan2025ApJ,Wei2025ApJ,Zhang2026ApJ}, analogous to the radiation in GRBs.

Currently, apart from a few candidate observations, such as ZTF19abanrhr \citep{Graham2020PRL} and GRB 191019A \citep{Levan2023NatAs}, no definitive observations of stellar-scale outburst events in AGN accretion disks have been reported. Given the scarcity of observational constraints, diverse radiation signatures predicted by different theoretical models cannot be ruled out, leaving considerable space for further theoretical development. Disk environments play an important role in structuring the signatures. Previous studies have mainly focused on disk models proposed by \cite{Sirko2003MNRAS} and \cite{Thompson2005ApJ}. The magnetically arrested disk (MAD) model has been extensively validated by numerical simulations (e.g., \citealp{Tchekhovskoy2011MNRAS}; \citealp{McKinney2012MNRAS,Narayan2012MNRAS,Aktar2026arXiv}), and is widely invoked to explain low-luminosity AGNs \citep[LLAGNs, e.g.,][]{Kadowaki2015ApJ,Scepi2021MNRAS,Chatterjee2025PRD}. For instance, recent Event Horizon Telescope observations of the centers of M87* and Sgr A* indicate that these objects harbor MADs \citep{Event2021ApJ,Event2024ApJ}. MADs form when sufficient magnetic flux accumulates in the inner disk to suppress accretion \citep{Narayan2003PASJ}, implying the presence of strong magnetic fields (MFs). If jets propagate in MAD environments, disk MFs may affect their dynamics and thereby potentially modify their observational signatures. \cite{Joshi2025AA} investigated the influence of strong MFs on accretion onto compact object mergers within magnetized AGN disks, showing that such fields can drive well-collimated outflows.

In this work, we investigate the effects of disk MFs on jet propagation and breakout emission in AGN MAD environments. The paper is organized as follows. In Section~2, we describe our model. In Section~3, we present our results. Conclusions and discussion are presented in Section~4.

\section{Method}
\subsection{MAD MF Strength}
MADs arise when poloidal MFs saturate near supermassive black holes (BHs) and become dynamically important, subsequently regulating mass accretion \citep{Narayan2003PASJ}. In this case, the MF strength at radius $R$ can be estimated by balancing the magnetic force associated with curved field lines against the gravitational force per unit area of disk matter in the radial direction, namely ${2 B_{\rm d}^2}/{4\pi} \sim {G M_{\rm BH}\Sigma}/{R^2}$, which yields \citep[e.g.,][]{Narayan2003PASJ}
\begin{equation}
B_{\rm d} \sim \sqrt{\frac{2 \pi G M_{\rm BH} \Sigma}{R^2}}.
\end{equation}
Here $G$ denotes the gravitational constant, $M_{\rm BH}$ is the mass of the supermassive BH, $R$ is the radial distance from the supermassive BH, and $\Sigma = 2 \rho_{\rm d} H_{\rm d}$ represents the surface density of the MAD\@. The quantities $\rho_{\rm d}$ and $H_{\rm d}$ denote the mass density and the vertical scale height, respectively.

For a geometrically thick, radiatively inefficient accretion flow characteristic of MADs, we adopt $H_{\rm d} = R/2$. The disk mass density can then be obtained from mass conservation,
\begin{equation}
\rho_{\rm d} = \frac{\dot M}{4\pi R H_{\rm d}\, v_R},
\end{equation}
where $\dot M$ is the mass accretion rate and $v_R$ is the radial inflow velocity. The accretion rate is parameterized as $\dot M = \dot m\, \dot M_{\rm Edd}$, where $\dot m$ is the dimensionless accretion rate and $\dot M_{\rm Edd}$ is the Eddington accretion rate, defined as $\dot M_{\rm Edd} = L_{\rm Edd}/(\eta_{\rm e} c^2)$. The Eddington luminosity is given by $L_{\rm Edd} = 4\pi G M_{\rm BH} m_{\rm p} c/\sigma_{\rm T}$. Similarly, the radial inflow velocity is expressed as a fraction of the free-fall velocity, $v_R = \epsilon\, v_{\rm ff}$, where $v_{\rm ff} = \sqrt{2 G M_{\rm BH}/R}$ is the free-fall velocity. Here $m_{\rm p}$ is the proton mass, $\sigma_{\rm T}$ is the Thomson cross section, $c$ is the speed of light, and $\eta_{\rm e}$ is the radiative efficiency. We adopt $\dot m = 1$, $\eta_{\rm e} = 0.1$, and $\epsilon = 0.01$ throughout this work \citep{Narayan2003PASJ,Narayan2014ARAA}.

\begin{figure*}[t]
\centering
\includegraphics[width=0.8\textwidth]{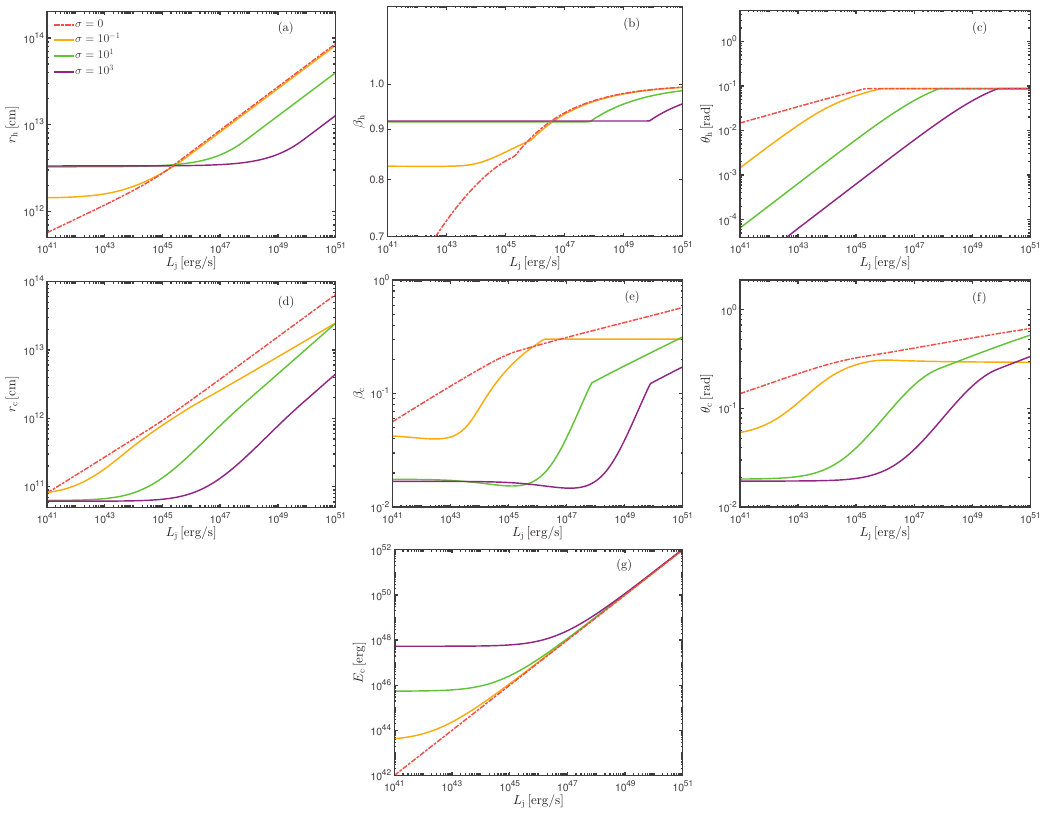}
\caption{Parameter evolution of a jet-cocoon system with jet power in a dense medium with a mass density of $10^{-11}\,\mathrm{g\,cm^{-3}}$, for different values of the generic ambient magnetization parameter $\sigma$. The red, yellow, green, and purple curves correspond to $\sigma = 0$, $10^{-1}$, $10^{1}$, and $10^{3}$, respectively. $\sigma$ is used only for an illustrative parameter survey. The physically realized MAD disk magnetization, denoted by $\sigma_{\rm d}$ in this work, is calculated self-consistently in Figure~\ref{fig2} and is typically $\sim10^{-3}\mbox{--}10^{-1}$. Thus, the $\sigma=10^{1}$ and $10^{3}$ cases should be regarded as illustrative high-magnetization limits.}
\label{fig1}
\end{figure*}

\subsection{Jet Dynamics}
When a powerful jet propagates through a dense environment, its interaction with the ambient medium leads to the formation of a hot cocoon surrounding the jet \citep[e.g.,][]{Zhang2003ApJ,Harrison2018MNRAS,Gottlieb2020MNRAS}. Thus, we adopt the jet-cocoon framework developed by \citet{Bromberg2011ApJ}, which builds upon earlier analytic treatments of cocoon pressure and jet propagation \citep{Begelman1989ApJ,Matzner2003MNRAS}, and incorporates disk MF effects to describe jet propagation within the MAD\@. In this framework, the jet head is described as a double-shock structure at the jet front, consisting of a forward shock propagating into the disk medium and a reverse shock propagating into the jet, separated by a contact discontinuity. We assume that the jet is launched from the MAD midplane and propagates perpendicular to the disk plane. We consider two potential channels through which the disk MF influences the jet-cocoon system. The first is the influence of magnetic pressure on the pressure balance at the jet head and at the cocoon sides, and the second is the contribution of magnetic energy dissipation, due to magnetic reconnection occurring on both sides of the cocoon, to the cocoon energy.

\subsubsection{Influence from Magnetic Pressure}
In the MAD, large-scale magnetic fields are predominantly poloidal (i.e., aligned with jet propagation) and do not provide head-on magnetic pressure to the jet head. However, potential small-scale turbulent fields or radial field components may be preferentially amplified by compression ahead of the jet head, exerting resistance to the motion of the jet head. Under these assumptions, with the jet itself taken to be nonmagnetized, the velocity of the jet head can be derived from pressure balance across the jet head, which reads \citep{Bromberg2011ApJ}
\begin{equation}
\rho_{\rm j} h_{\rm j} c^2 \Gamma_{\rm j}^2 \Gamma_{\rm h}^2 (\beta_{\rm j} - \beta_{\rm h})^2 + P_{\rm j}=\rho_{\rm d} h_{\rm d,eff} c^2 \Gamma_{\rm h}^2 \beta_{\rm h}^2 + P_{\rm th,d} + P_{B,\rm h},
\label{eq3}
\end{equation}
where $\rho$, $P$, $\Gamma$, and $\beta$ denote the mass density, pressure, Lorentz factor, and dimensionless velocity of each fluid, respectively. Subscripts ${\rm j}$, ${\rm h}$, and ${\rm d}$ refer to the jet, the jet head, and the disk medium, respectively. We parameterize the magnetic pressure contributed by the field component transverse to the jet-head motion as $P_{B,\rm h} = f_{\rm h} B_{\rm d}^2 / 8\pi$, where $0 \leq f_{\rm h} \leq 1$. The dimensionless specific enthalpy of the jet is denoted by $h_{\rm j}$. The effective specific enthalpy of the disk medium upstream of the jet head is given by
\begin{equation}
h_{\rm d,eff} = h_{\rm d}\left(1 + \sigma_{\rm h}\right),
\end{equation}
where $h_{\rm d} = 1 + 4P_{\rm th,d}/(\rho_{\rm d} c^2)$ and $\sigma_{\rm h} = f_{\rm h}\sigma_{\rm d}$. In this definition, $\sigma_{\rm d} = B_{\rm d}^2/(4\pi\rho_{\rm d}h_{\rm d}c^2)$ represents the total magnetization of the disk, while $\sigma_{\rm h}$ describes the effective magnetization at the jet head arising exclusively from the transverse magnetic field component. The nonmagnetized-jet assumption can be understood as a hydrodynamic approximation. Namely, even though the jet is magnetically launched near its compact-object engine, we assume that most of its Poynting flux has been converted into kinetic energy before the jet-head balance becomes applicable. Following previous studies \citep[e.g.,][]{Bromberg2011ApJ}, we assume that a strong reverse shock forms at the jet head and that the temperature of the disk medium remains nonrelativistic, and therefore the thermal pressures $P_{\rm j}$ and $P_{\rm th,d}$ are negligible compared to the corresponding ram-pressure terms. Under these assumptions, the jet-head velocity, $\beta_{\rm h}$, is calculated numerically.

\begin{figure*}[t]
\centering
\includegraphics[width=0.315\textwidth]{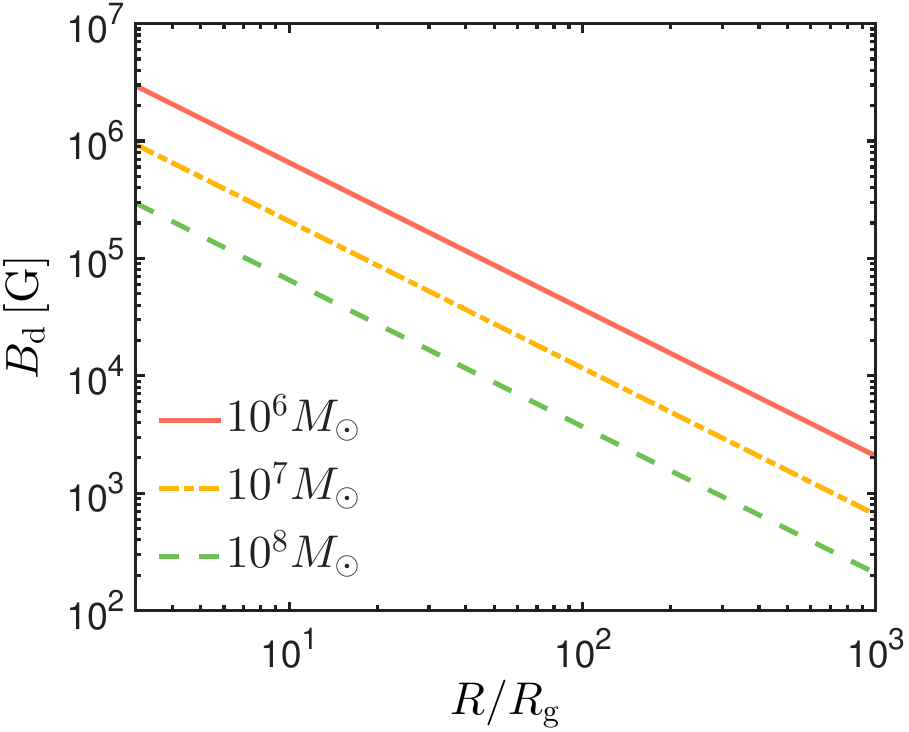}\hfill
\includegraphics[width=0.315\textwidth]{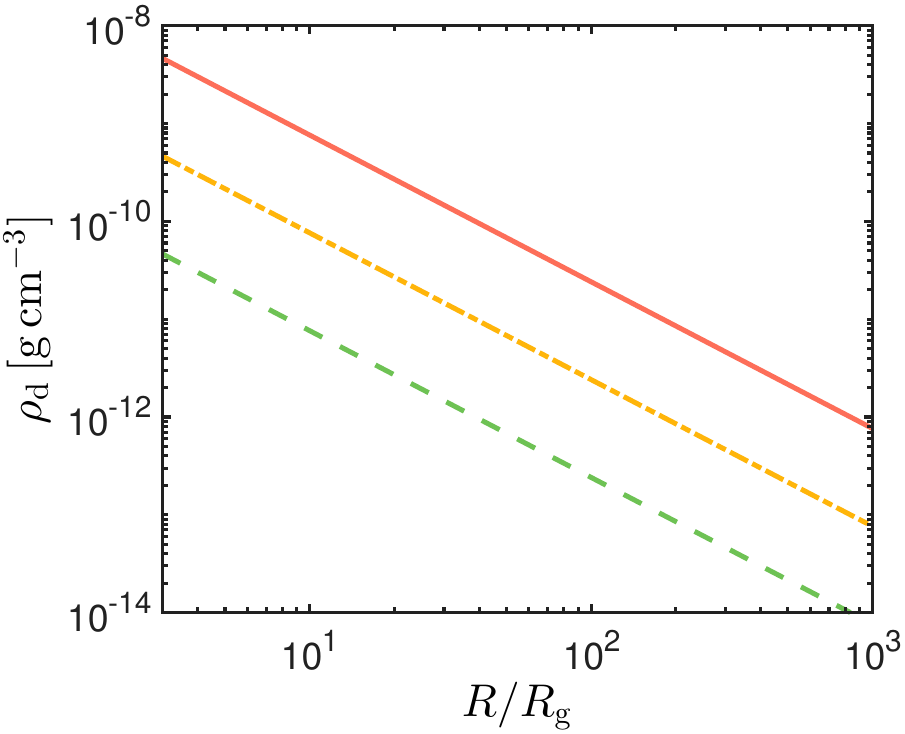}\hfill
\includegraphics[width=0.315\textwidth]{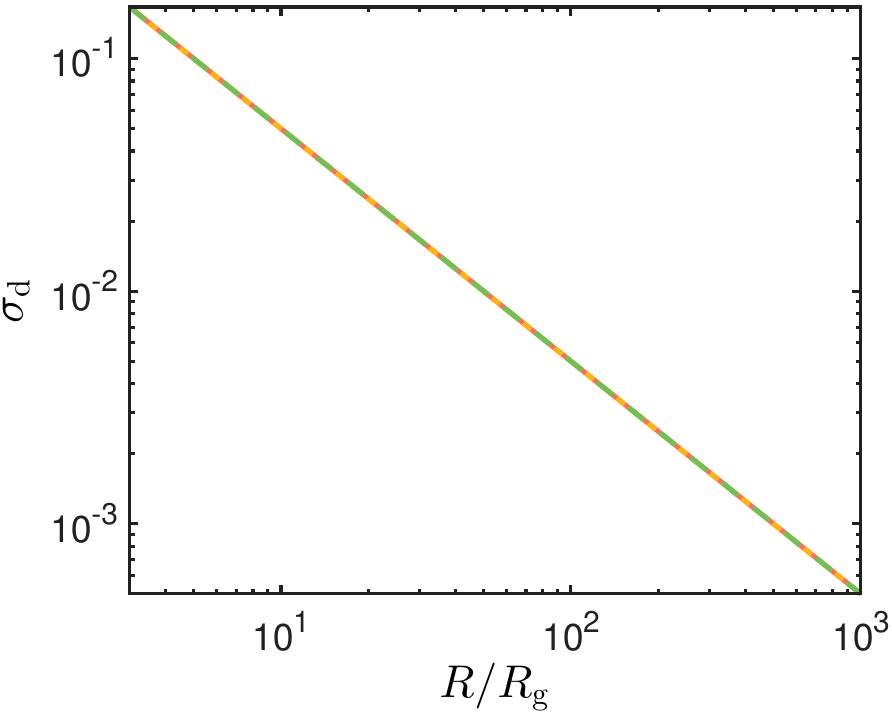}
\caption{Radial profiles of the MF strength, mass density, and magnetization parameter in MADs. The green dashed, yellow dot-dashed, and red solid lines correspond to supermassive BH masses $M_{\rm BH} = 10^{6}$, $10^{7}$, and $10^{8}\,M_{\odot}$, respectively.}
\label{fig2}
\end{figure*}

On the other hand, for the lateral expansion of the cocoon, the poloidal field naturally generates a substantial magnetic pressure $P_{B,\rm c} = f_{\rm c} B_{\rm d}^2 / 8\pi$, where $f_{\rm c} \approx 1$, which opposes radial motion of the cocoon. The assumption $f_{\rm c}\simeq1$ represents a fiducial approximation for a coherent poloidal field. However, it should be noted that imperfect field coherence, turbulence, or a substantial nonpoloidal component could reduce the effective lateral magnetic pressure and weaken the predicted cocoon confinement. Accounting for magnetic pressure effects, the lateral expansion velocity of the cocoon is given by
\begin{equation}
\beta_{\rm c} = \min \left( \sqrt{\frac{P_{\rm c}-P_{B,\rm c}}{\bar{\rho}_{\rm d}\,c^2}}, \beta_{\rm A} \right),
\label{eq5}
\end{equation}
where $P_{\rm c}$, $\bar{\rho}_{\rm d}$, and $\beta_{\rm A} = \sqrt{\frac{\sigma_{\rm d}}{1+\sigma_{\rm d}}}$ denote the cocoon pressure, the average disk density, and the dimensionless Alfv\'en speed, respectively \citep{Bromberg2011ApJ}. When $P_{\rm c}<P_{B,\rm c}$, the cocoon pressure is insufficient to overcome the magnetic pressure exerted by the disk field. In this case, the lateral expansion of the cocoon is magnetically confined; we therefore set $\beta_{\rm c}=0$.

We note that if the field lines are bent around the jet head or the cocoon boundary, magnetic tension may also contribute to the force on the jet-cocoon system. The magnitude of this contribution depends on the local field-line curvature and on the geometry of the magnetic draping layer, which cannot be determined self-consistently in the present analytic model. We therefore do not include a separate magnetic-tension term. Instead, its possible effect is incorporated phenomenologically into the geometrical factors \(f_{\rm h}\) and \(f_{\rm c}\).

\subsubsection{Influence from Magnetic Energy Dissipation}
As the central engine continues to operate, providing the jet with energy, both the position of the jet head and the lateral radius of the cocoon evolve significantly with time. $z_{\rm h}$ denotes the vertical position of the jet head measured from the disk midplane, and $r_{\rm c}$ denotes the lateral radius of the cocoon. Their temporal evolution can be described by \citep{Bromberg2011ApJ,Chen2025ApJ}
\begin{equation}
\frac{d z_{\rm h}}{d t} = \beta_{\rm h} c,
\end{equation}
and
\begin{equation}
\frac{d r_{\rm c}}{d t} = \beta_{\rm c} c,
\end{equation}
respectively. In addition, the time evolution of the cocoon internal energy is given by
\begin{equation}
\frac{d E_{\rm c}}{d t} = \eta_{\rm h} L_{\rm j} \left( 1 - \beta_{\rm h} \right) + L_{B},
\end{equation}
where $\eta_{\rm h} = \min \left( 2/(\Gamma_{\rm h}\theta_{\rm h}),\, 1 \right)$ denotes the fraction of the jet-head energy that is deposited into the cocoon, with $\Gamma_{\rm h} = (1-\beta_{\rm h}^2)^{-1/2}$ being the Lorentz factor of the jet head, and $\theta_{\rm h} = \arctan \left( r_{\rm h}/z_{\rm h} \right)$ being its half-opening angle. Here, $r_{\rm h}$ is the cross-sectional radius of the jet head, defined as $r_{\rm h} = \min (z_{\rm h},\sqrt{(3 L_{\rm j} z_{\rm h} r_{\rm c}^2)/(4 c E_{\rm c})}) \tan \theta_{\rm j}$, where $\theta_{\rm j}$ is the half-opening angle of the initial jet. Similarly, the half-opening angle of the cocoon is defined as $\theta_{\rm c} = \arctan ( r_{\rm c}/z_{\rm h})$. $L_{\rm j}$ is the jet power from the central engine and $L_{B}$ is the power of magnetic energy dissipation.
We adopt $\Gamma_{\rm j} = 100$ and $\theta_{\rm j} = 5^\circ$ throughout this work \citep{Zhang2019book}.

For magnetic energy dissipation, we consider the following scenario. As the jet-cocoon system propagates within the MAD, abundant small-scale turbulence is produced at the lateral boundary of the cocoon due to the shear motions occurring between the system and the disk medium. This turbulence in turn triggers fast magnetic reconnection, which dissipates magnetic energy. The released magnetic energy is injected into the cocoon, thereby increasing its internal energy. In this scenario, the power associated with turbulence-driven magnetic reconnection can be written as \citep[e.g.,][]{Gouveia2005AA,Gouveia2010AA,Kadowaki2015ApJ}
\begin{equation}
{L}_{B} = \frac{B_{\rm d}^{2}}{8\pi}\, v_{\rm A}\,\left( 2\pi r_{\rm c}\,\Delta R_{\rm x} \right),
\end{equation}
where $v_{\rm A} = \beta_{\rm A}c$ is the Alfv\'en speed and $\Delta R_{\rm x}$ is the width of the current sheet, which can be obtained from \citep[e.g.,][]{Lazarian1999ApJ,Kadowaki2015ApJ}
\begin{equation}
\Delta R_{\rm x} = z_{\rm x} \min \left[ \frac{z_{\rm inj}}{z_{\rm x}}, \, \frac{z_{\rm x}}{z_{\rm inj}} \right]^{1/2} M_{\rm A}^2.
\end{equation}
Here, $z_{\rm x}$ denotes the spatial extent of the reconnection region, and $M_{\rm A} \equiv v_{\rm inj}/v_{\rm A}$ is the Alfv\'enic Mach number of the turbulence, where $v_{\rm inj}$ is the injected turbulent velocity at the injection scale $z_{\rm inj}$. For turbulence driven by shear motions, the injected turbulent velocity is typically $\sim 0.1\mbox{--}0.3$ times the shear velocity \citep[e.g.,][]{Dimotakis2000JFM,Dimotakis2005ARFM}. In the present scenario, the disk medium is approximately stationary with respect to the vertically propagating jet-cocoon system. Consequently, the shear velocity can be approximated by the velocity of the jet head, $v_{\rm h}$. We therefore adopt $v_{\rm inj} = \min \left( 0.3 v_{\rm h}, v_{\rm A} \right)$. Furthermore, since the global magnetic-reconnection region can be regarded as comprising a large number of smaller reconnection regions, we assume $z_{\rm inj} = z_{\rm x}$ and set the reconnection scale to $z_{\rm x} = z_{\rm h}$. It should be noted that the choice $z_{\rm x} = z_{\rm h}$ corresponds to an idealized, maximal reconnection scale.

\begin{figure}[t]
\centering
\includegraphics[width=0.80\columnwidth]{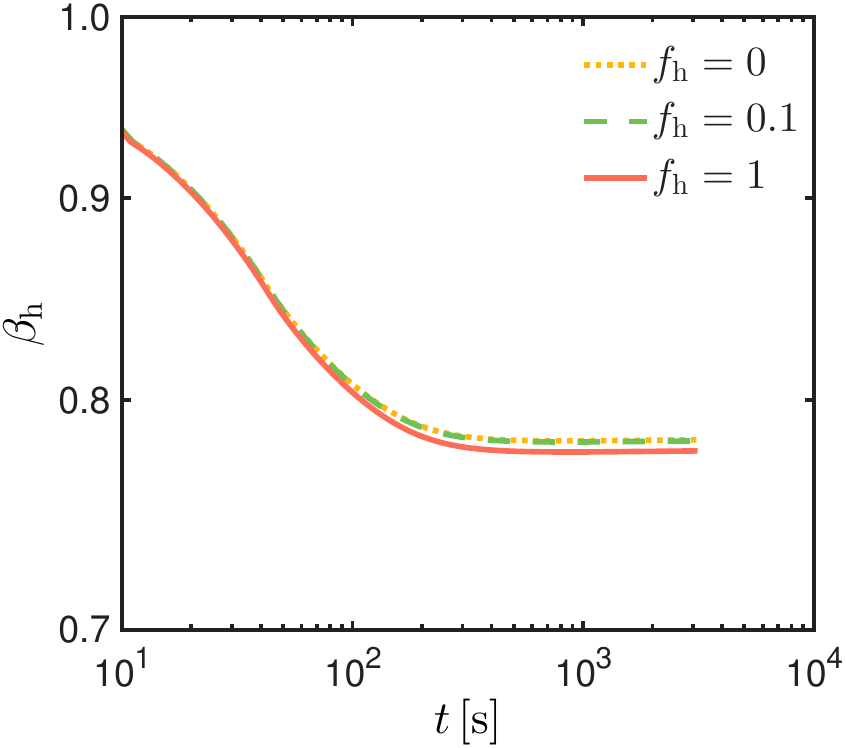}
\caption{Temporal evolution of the jet-head velocity for different values of \(f_{\rm h}\). The yellow dotted, green dashed, and red solid curves correspond to \(f_{\rm h}=0.0\), \(0.1\), and \(1\), respectively.}
\label{fig3}
\end{figure}

\subsection{Thermal Emission from Jet Breakout}
The jet head remains optically thick while propagating through the dense environment of the MAD\@. As a result, the radiation generated in the shocked region is expected to be thermalized. In addition, since the timescale of photon diffusion to the disk surface exceeds the dynamical timescale of jet-head propagation at early times, the radiation remains trapped behind the forward shock and cannot escape. Breakout occurs when the photon diffusion timescale becomes comparable to the dynamical timescale, i.e., $t_{\rm diff} \sim t_{\rm dyn}$. We therefore define the breakout location of the jet-head shock, $z_{\rm h,b}$, by equating these two timescales, such that $\tau(z_{\rm h,b}) = c/v_{\rm h,b}$, where $\tau(z_{\rm h,b})$ is the optical depth of the disk medium measured from $z_{\rm h,b}$ to the disk surface and $v_{\rm h,b}$ is the jet-head velocity at the breakout location. We treat the jet-head velocity obtained from the pressure balance as an approximation to the forward-shock velocity when evaluating the breakout emission. We consider only cases of successful jet breakout and assume that the engine duration is sufficiently long for the jet-head shock to reach the breakout location before the jet base catches up with the head. We calculate the breakout luminosity and the corresponding characteristic temperature.

When the jet head is Newtonian, the breakout luminosity of the jet-head shock can be estimated from the kinetic energy flux across the shock surface \citep{Nakar2010ApJ}
\begin{equation}
L_{\rm h,n} = \Sigma_{\rm h,b}\,\rho_{\rm d,b}\,\beta_{\rm h,b}^{3}\,c^{3},
\end{equation}
where $\Sigma_{\rm h,b}$ and $\beta_{\rm h,b}$ are the cross-sectional area and velocity of the jet head at the breakout location, and $\rho_{\rm d,b}$ is the disk mass density at the breakout location. The subscript ${\rm b}$ denotes quantities evaluated at breakout. For a sufficiently slow head ($\beta_{\rm h,b} \leq 0.03$), the post-shock radiation has sufficient time to reach thermodynamic equilibrium. The breakout temperature is therefore given by \citep{Sapir2011ApJ}
\begin{equation}
T_{\rm h,n}=\left(\frac{18}{7a}\,\rho_{\rm d,b}\,\beta_{\rm h,b}^{2}\,c^{2}\right)^{1/4},
\end{equation}
where $a$ is the radiation constant. For a jet head with $0.03 < \beta_{\rm h,b} < 0.4$, the radiation field is generally out of full thermal equilibrium. The photon spectrum is modified by Comptonization, and the breakout temperature can be estimated as \citep{Sapir2013ApJ}
\begin{equation}
\begin{aligned}
\log_{10}\!\left(\frac{k_{\rm B}T_{\rm h,Comp}}{\mathrm{eV}}\right)
&= 0.975 + 1.735\left(\frac{\beta_{\rm h,b}}{0.1}\right)^{1/2} \\
&\quad + \left[
0.26 - 0.08\left(\frac{\beta_{\rm h,b}}{0.1}\right)^{1/2}
\right] \\
&\quad \times \log_{10}\!\left(
\frac{n_{\rm b}}{10^{15}\,\mathrm{cm^{-3}}}
\right),
\end{aligned}
\end{equation}
where $n_{\rm b} = \rho_{\rm d,b}/m_{\rm p}$ is the number density of the disk medium at the breakout location, $k_{\rm B}$ is the Boltzmann constant, and $T_{\rm h,Comp}$ is the Comptonized breakout temperature.

\begin{figure*}[t]
\centering
\includegraphics[width=0.8\textwidth]{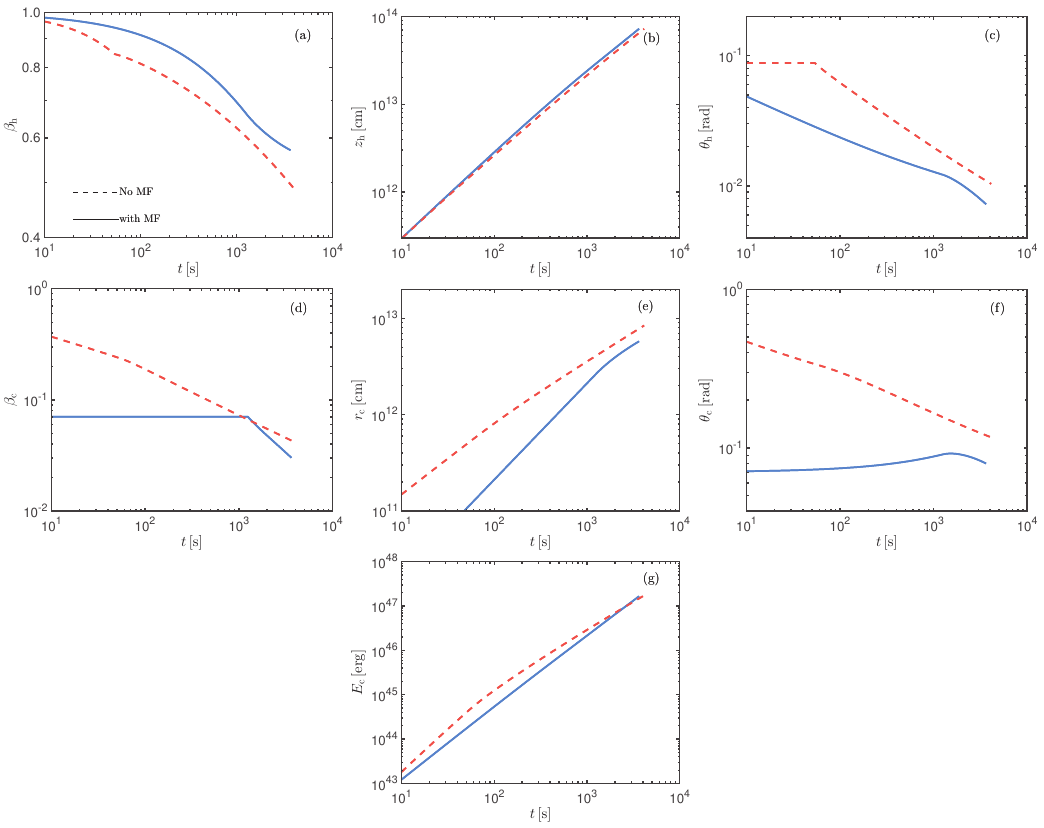}
\caption{Parameter evolution of jet-cocoon systems in a MAD with $M_{\rm BH} = 10^{7}\,M_{\odot}$. The jet is launched from a radial location of $R = 10^{2}\,R_{\rm g}$, with Lorentz factor $\Gamma_{\rm j} = 100$ and initial half-opening angle $\theta_{\rm j} = 5^\circ$. Jet power is fixed at $L_{\rm j} = 10^{44}\,\mathrm{erg/s}$. Dashed lines denote the results obtained without MF effects. The evolution is terminated at the breakout time of jet-head shocks.}
\label{fig4}
\end{figure*}

In the relativistic regime ($\beta_{\rm h,b} \geq 0.5$), the emission characteristics differ significantly from the above cases due to relativistic shock structure and radiative transfer effects. The breakout luminosity can be approximated as \citep{Chen2025ApJ}
\begin{equation}
L_{\rm h,r}=\frac{\gamma_{\rm h,th}^{3}}{2\gamma_{\rm h}}\frac{E_0}{t_{\rm h,th}},
\end{equation}
where $\gamma_{\rm h}$ is the Lorentz factor of the shocked fluid in the jet head, $E_0$ is the internal energy of the breakout shell, $t_{\rm h,th}$ is the time at which the shell becomes optically thin, and $\gamma_{\rm h,th}$ is the Lorentz factor at transparency. In addition, the breakout temperature is \citep[e.g.,][]{Nakar2012ApJ,Chen2025ApJ}
\begin{equation}
T_{\rm h,r}=\gamma_{\rm h,th}\,T'_{\rm th},
\end{equation}
where $k_{\rm B}T'_{\rm th} = 50\,\mathrm{keV}$, with $T'_{\rm th}$ being the comoving temperature of the shell at transparency. Following \citet{Chen2025ApJ}, we extrapolate the relativistic-shell prescription down to $\beta_{\rm h,b}>0.4$. Over the intervening interval $0.4<\beta_{\rm h,b}<0.5$, we perform a smooth interpolation of the breakout luminosity and temperature.

By analogy with the Newtonian jet-head case, we roughly estimate the cocoon luminosity at the breakout location of the jet-head shock as
\begin{equation}
L_{\rm c}
=
\pi r_{\rm c,b}^{2}\,\rho_{\rm d,b}\,\beta_{\rm c,b}^{3}\,c^{3},
\end{equation}
where $\beta_{\rm c,b}$ is the cocoon velocity at the breakout location.

\begin{figure*}[t]
\centering
\includegraphics[width=0.8\textwidth]{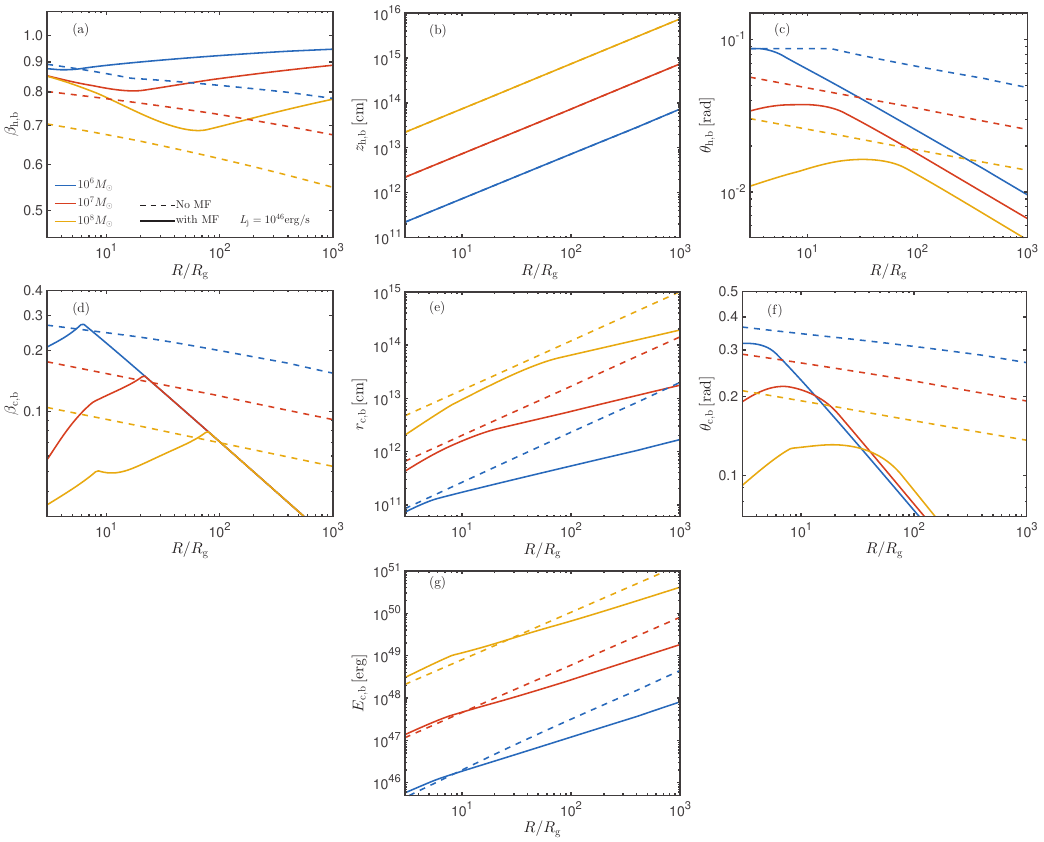}
\caption{Radial distributions of jet-cocoon parameters evaluated at the breakout location of jet-head shocks. The blue, red, and yellow lines denote the results obtained with $M_{\rm BH} = 10^{6}$, $10^{7}$, and $10^{8}\,M_{\odot}$, respectively. Solid and dashed lines denote the results obtained with and without MF effects, respectively. The jet power is fixed at $L_{\rm j} = 10^{46}\,\mathrm{erg/s}$.}
\label{fig5}
\end{figure*}

\section{Results}
We first compute the dynamics of jet-cocoon systems propagating within dense environments as functions of jet power for several magnetization levels, namely $\sigma = 0$, $10^{-1}$, $10^{1}$, and $10^{3}$. We consider a broad range of jet power, $L_{\rm j} \sim 10^{41}\mbox{--}10^{51}\,\mathrm{erg/s}$, assuming an ambient density of $10^{-11}\,\mathrm{g\,cm^{-3}}$ and adopting $f_{\rm h}=0.1$. For each jet power, the dynamics are calculated at the time when the jet base catches up with the jet head. This timescale is governed by $t_{\rm en} = t_{\rm j} + \int_{0}^{t_{\rm en}} \beta_{\rm h}\, dt$, where $t_{\rm j}$ denotes the central engine activity duration, set to $10\,\mathrm{s}$ \citep{Piran2004RMP}. Here, \(t_{\rm j}=10\,\mathrm{s}\) is adopted only for illustrative dynamical calculations in Figure~\ref{fig1}; otherwise, we assume an engine duration or sustained energy supply sufficient for successful jet breakout. As shown in Figure~\ref{fig1}, the impact of ambient MFs on jet-cocoon dynamics strengthens with increasing magnetization. Furthermore, a higher magnetization corresponds to a broader jet-power range over which jet-cocoon dynamics are shaped by the MFs. It should be noted that Figure~\ref{fig1} serves as an illustrative parameter survey of generic ambient magnetization, demonstrating how jet-cocoon dynamics respond across magnetized environments ranging from weak to strong.

\subsection{MAD Magnetization Parameter}
Figure~\ref{fig2} presents the magnetic-field strength (left panel), mass density (middle panel), and magnetization parameter (right panel) within MADs, evaluated over radial distances spanning $3\mbox{--}10^3\,R_{\rm g}$, for several supermassive BH masses. Here, $R_{\rm g}$ is the gravitational radius of the supermassive BH, defined as $R_{\rm g} \equiv G M_{\rm BH}/c^2$. In the left panel, the MF strength spans $\sim10^{3}\mbox{--}10^{6}\,\mathrm{G}$. It decreases with increasing radius and is inversely correlated with the BH mass. The middle panel indicates that the mass density ranges from approximately $10^{-13}$ to $10^{-9}\,\mathrm{g\,cm^{-3}}$, declining gradually with radius.
Moreover, the right panel indicates that the magnetization parameter lies in the range of $\sim10^{-3}\mbox{--}10^{-1}$. We note that the disk magnetization $\sigma_{\rm d}$, representing the self-consistently computed MAD disk magnetization, should be distinguished from the generic ambient magnetization parameter $\sigma$, which is adopted solely for the illustrative parameter survey in Figure~\ref{fig1}. In addition, the $\sigma=10^{1}$ and $10^{3}$ cases in Figure~\ref{fig1} should be regarded as illustrative high-magnetization limits rather than fiducial MAD values. When the ambient medium is identified with the MAD disk, we have $\sigma_{\rm d} = \sigma$.

\subsection{Effects of Factor \texorpdfstring{$f_{\rm h}$}{fh}}
\label{sub:EF}

\begin{figure*}[t]
\centering
\includegraphics[width=0.8\textwidth]{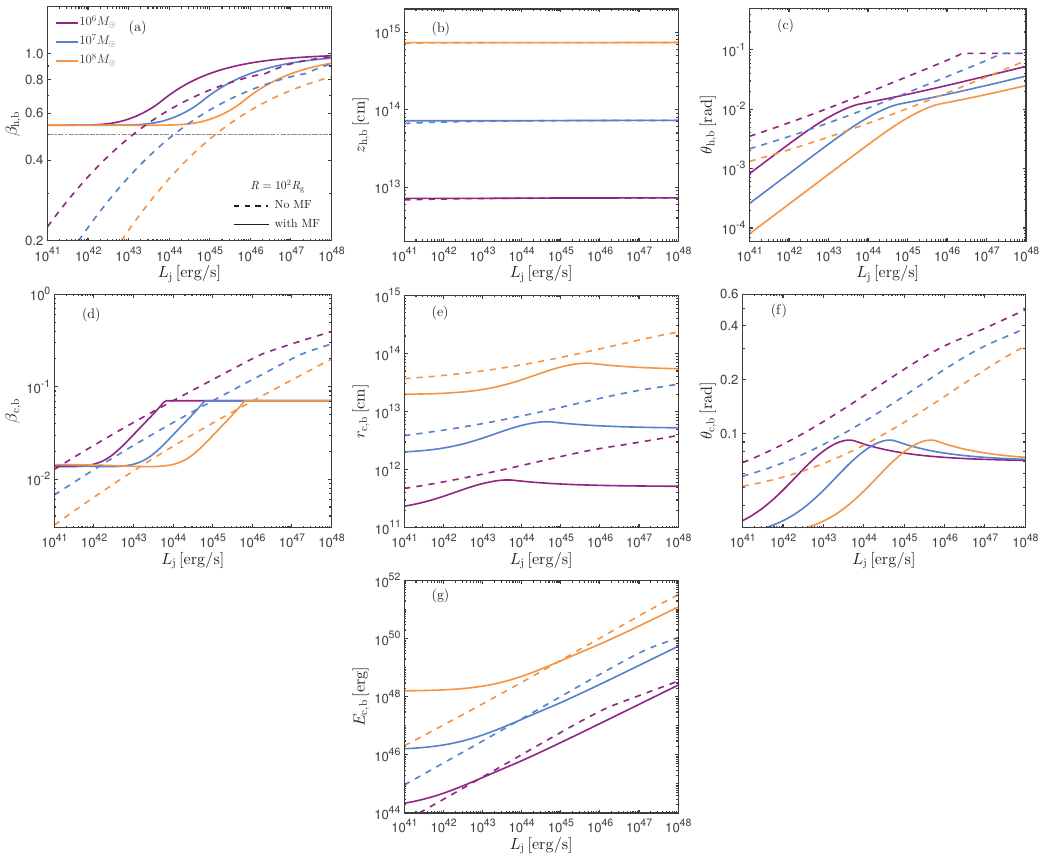}
\caption{Dynamical parameters of jet-cocoon systems at the breakout location as functions of jet power. The purple, blue, and yellow lines denote the results obtained with $M_{\rm BH} = 10^{6}$, $10^{7}$, and $10^{8}\,M_{\odot}$, respectively. Solid and dashed lines denote the results obtained with and without MF effects, respectively. The gray dash-dotted line in panel~(a) marks $\beta_{\rm h,b}=0.5$, and the $\beta_{\rm h,b}=0.03$ boundary lies outside the plotted range. The radial location of jet launching is fixed at $R = 10^{2}\,R_{\rm g}$.}
\label{fig6}
\end{figure*}

To examine the dynamical effect of uncertain magnetic-field geometries upstream of the jet head, we have parameterized the head-on magnetic pressure as $P_{B,\rm h}=f_{\rm h}{B_{\rm d}^2}/{8\pi}$. We consider three representative cases, $f_{\rm h}=0$, $0.1$, and $1$. The $f_{\rm h}=0$ case corresponds to purely large-scale poloidal fields, $f_{\rm h}=0.1$ represents a modest transverse component which may be produced by turbulence or partial magnetic draping, and $f_{\rm h}=1$ gives an upper limit, assuming full reorientation of large-scale poloidal fields into transverse components ahead of the jet head. We calculate the temporal evolution of the jet-head velocity for these three cases. To maximize the dynamical impact of $f_{\rm h}$, we employ a fiducial parameter set characterized by a relatively strong magnetic field and a low jet power, i.e., $M_{\rm BH}=10^{8}\,M_{\odot}$, $R=10\,R_{\rm g}$, $\Gamma_{\rm j}=100$, $\theta_{\rm j}=5^\circ$, and $L_{\rm j}=10^{44}\,\mathrm{erg/s}$. As shown in Figure~\ref{fig3}, the curves for $f_{\rm h}=0$ and $0.1$ are nearly indistinguishable, while the $f_{\rm h}=1$ case shows only a slight reduction. This weak dependence on $f_{\rm h}$ arises because the ram pressure of the disk medium remains much larger than the head-on magnetic pressure under the present conditions. A simple estimate illustrates this point. For $f_{\rm h}=0.1$, the disk ram pressure is $P_{\rm ram,d}=\rho_{\rm d} h_{\rm d,eff} c^2 \Gamma_{\rm h}^2 \beta_{\rm h}^2 \sim 1.22\times10^{10}\,\mathrm{erg\,cm^{-3}}$, whereas the head-on magnetic pressure is $P_{B,\rm h} \sim 1.68\times10^{7}\,\mathrm{erg\,cm^{-3}}$. This gives $P_{B,\rm h}/P_{\rm ram,d}\simeq1.4\times10^{-3}$. Even in the upper-limit case $f_{\rm h}=1$, we find $P_{\rm ram,d} \sim 1.28\times10^{10}\,\mathrm{erg\,cm^{-3}}$ and $P_{B,\rm h} \sim 1.68\times10^{8}\,\mathrm{erg\,cm^{-3}}$, corresponding to $P_{B,\rm h}/P_{\rm ram,d}\simeq1.3\times10^{-2}$. In addition, the effective head magnetization is also small, $\sigma_{\rm h}\simeq4.9\times10^{-3}$ for $f_{\rm h}=0.1$, and $\sigma_{\rm h}\sim4.9\times10^{-2}$ for $f_{\rm h}=1$. These results indicate that, for the jet head propagating through the dense MAD medium, the direct magnetic terms at the head are dynamically subdominant. We therefore adopt \(f_{\rm h}=0.1\) for all subsequent calculations.

\subsection{Jet-Cocoon System Evolution}
\subsubsection{Time Evolution}
Figure~\ref{fig4} shows the temporal evolution of the dynamical parameters of the jet-cocoon system. We adopt a supermassive BH with a mass of $M_{\rm BH} = 10^{7}\,M_{\odot}$. In addition, to better illustrate how the MF influences the temporal evolution of jet-cocoon dynamics, we employ a relatively low jet power, $L_{\rm j} = 10^{44}\,\mathrm{erg/s}$. The solid and dashed lines denote the cases with and without MF effects, respectively.

For the jet head, the jet-head velocity exhibits a modest, collimation-driven increase when MF effects are included, whereas the half-opening angle is substantially reduced, implying enhanced jet collimation as the jet propagates with time. For the cocoon, the lateral expansion velocity evolves more slowly at early times when MF effects are included, owing to the constraint imposed by the dimensionless Alfv\'en speed (see Equation~\eqref{eq5}). Correspondingly, both the lateral expansion radius and the half-opening angle are substantially reduced, indicating that magnetic pressure effectively suppresses cocoon expansion. Meanwhile, the cocoon energy exhibits a mild decrease in the case with MF effects relative to the nonmagnetized case. This behavior arises because the magnetic dissipation power at early epochs remains weaker than the power injected into the cocoon by the jet. The cocoon internal energy is therefore dominated by the jet power, and the fraction of jet power deposited within the cocoon declines as the jet-head velocity increases.

\subsubsection{Radial Distribution}
Figure~\ref{fig5} presents the radial evolution of the dynamical parameters of the jet-cocoon system over $R=3\mbox{--}10^{3}\,R_{\rm g}$ at the breakout location of jet-head shocks. $\beta_{\rm h,b}$, $z_{\rm h,b}$, and $\theta_{\rm h,b}$ denote the dimensionless velocity, location, and half-opening angle of the jet head at breakout, respectively; the corresponding cocoon quantities are $\beta_{\rm c,b}$, $r_{\rm c,b}$, and $\theta_{\rm c,b}$. $E_{\rm c,b}$ denotes the cocoon energy at the breakout location. We fix the jet power to $L_{\rm j} = 10^{46}\,\mathrm{erg/s}$.

As the radial distance increases, the suppressive influence of MFs on dynamical parameters becomes more pronounced, with the jet-head velocity and location constituting notable exceptions. In panel~(a), the velocity first decreases with increasing radius in the case with MF effects. However, this trend reverses beyond a certain location. This transition originates from the rapid decline of the jet-head half-opening angle with increasing radius, as illustrated in panel~(c). The smaller half-opening angle leads to a higher jet density and consequently produces a modest increase in the jet-head velocity. In panel~(b), the jet-head location is nearly identical with and without MF effects; notably, this feature exhibits little radial dependence. In panel~(c), the half-opening angle is smaller in the case with MF effects than in the nonmagnetized case, and this difference grows progressively more pronounced at larger radii, implying stronger cocoon confinement with increasing radial distance. This behavior arises because the disk scale height increases with radius; thus, the jet head takes longer to reach the breakout location, allowing the confinement effect to accumulate.

In panel~(d), the cocoon expansion velocity initially increases with radial distance but subsequently decreases. This behavior can be understood as follows: as the radial distance increases, the magnetic pressure becomes relatively weaker, thereby allowing the expansion velocity to increase at first. However, once the velocity approaches the Alfv\'en speed, the expansion becomes regulated by it. Since the Alfv\'en speed also decreases with the declining MF strength, the cocoon expansion velocity decreases at larger radii. In panel~(e), the cocoon radius is suppressed by large-scale MFs, and this suppression grows markedly as the radial distance increases. A similar trend is found for the cocoon half-opening angle in panel~(f). This behavior originates from the fact that the cocoon is subject to magnetic compression over a longer duration at larger radial distances, and as a result, the cumulative MF influence becomes progressively stronger. Finally, in panel~(g), the cocoon energy in the case with MF effects is initially slightly higher than that in the nonmagnetized case, primarily due to the contribution from magnetic energy dissipation. However, as the radial distance increases, the cocoon energy in the case with MF effects becomes lower than that in the nonmagnetized case. This transition occurs because MF strength declines with radial distance, reducing the total dissipated magnetic energy. Nevertheless, cocoon confinement still produces a modest, collimation-driven increase in the jet-head velocity, lowering the fraction of jet energy deposited within the cocoon. Consequently, the cocoon receives comparatively less jet energy in the case with MF effects.

\subsubsection{Evolution with Jet Power}

\begin{figure}[t]
\centering
\includegraphics[width=0.80\columnwidth]{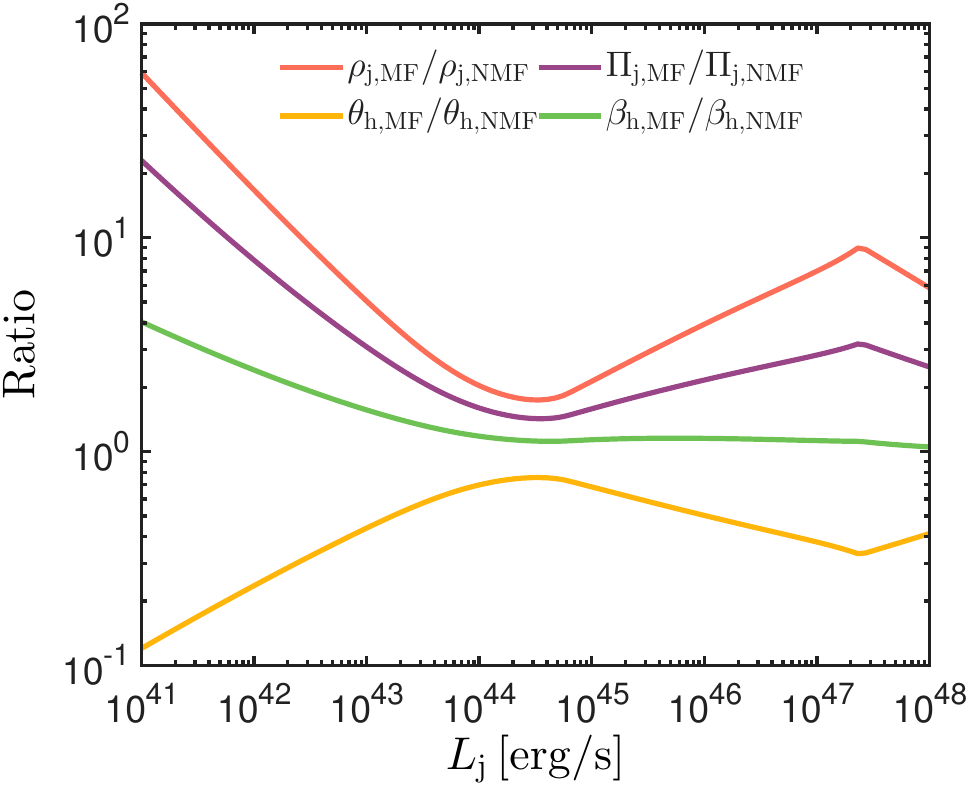}
\caption{Ratios of dynamical parameters between the magnetized and nonmagnetized cases as functions of jet power. The red, purple, yellow, and green curves show $\rho_{\rm j,MF}/\rho_{\rm j,NMF}$, $\Pi_{\rm j,MF}/\Pi_{\rm j,NMF}$, $\theta_{\rm h,MF}/\theta_{\rm h,NMF}$, and $\beta_{\rm h,MF}/\beta_{\rm h,NMF}$, respectively.}
\label{fig7}
\end{figure}

\begin{figure*}[t]
\centering
\includegraphics[width=0.4\linewidth, height=0.3\linewidth]{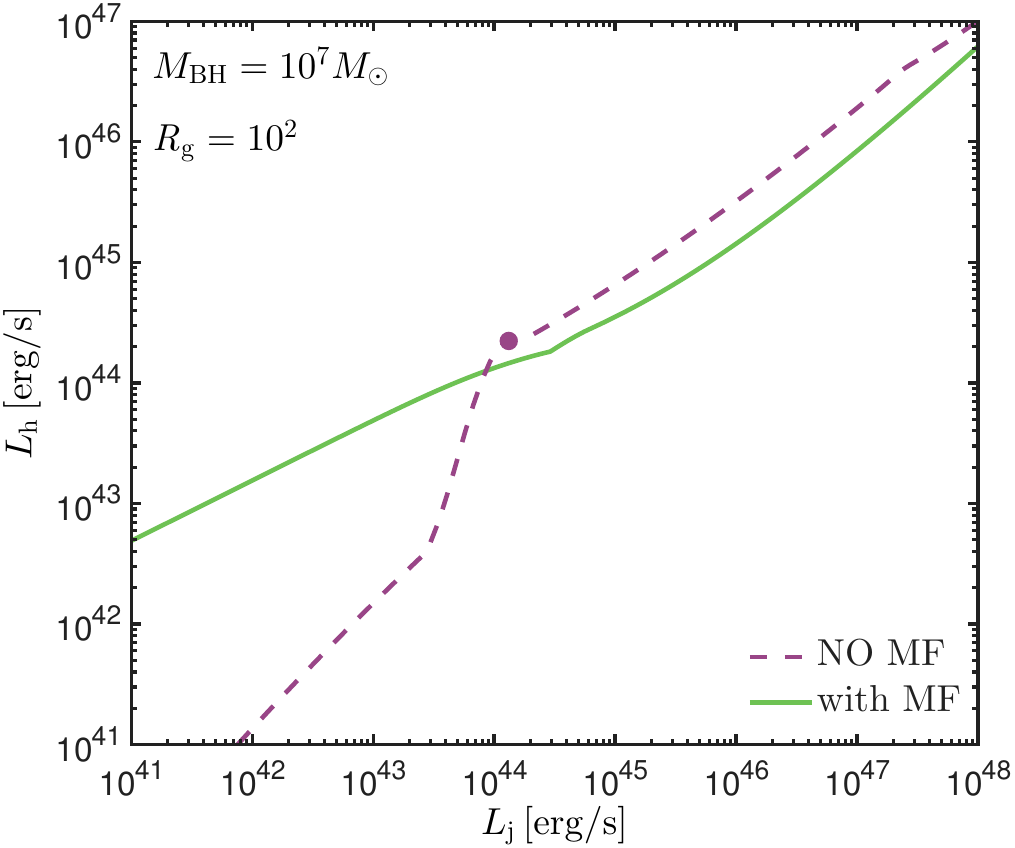}
\includegraphics[width=0.4\linewidth, height=0.3\linewidth]{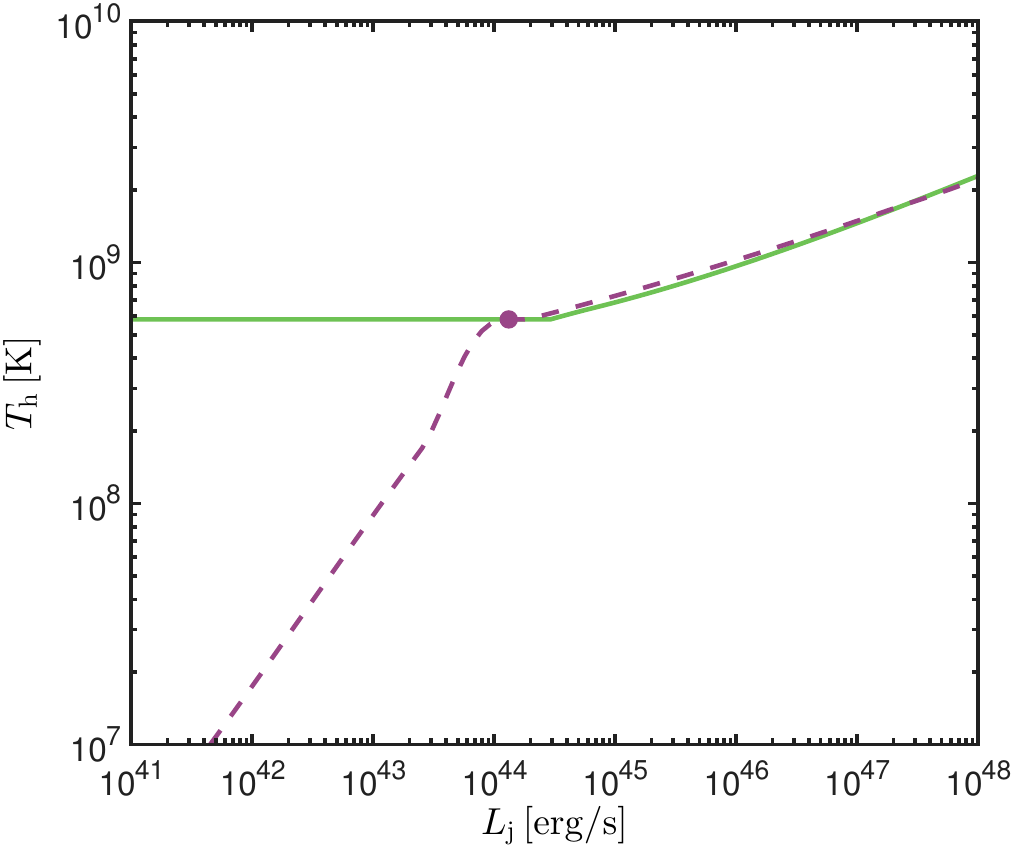}
\caption{Luminosity and temperature of the jet-head shock breakout as functions of jet power in a MAD for $M_{\rm BH} = 10^{7}\,M_{\odot}$. The left and right panels show the breakout luminosity and breakout temperature, respectively. Solid and dashed lines represent results obtained with and without MF effects, respectively. The solid purple dots mark $\beta_{\rm h,b}=0.5$ on the nonmagnetized curves, above which the fully relativistic prescription is applied; all magnetized solutions shown here already satisfy $\beta_{\rm h,b}>0.5$. The jet is launched at a fixed radius of $R = 10^{2}\,R_{\rm g}$.}
\label{fig8}
\end{figure*}

Figure~\ref{fig6} shows the jet-cocoon dynamics at the breakout location of jet-head shocks as a function of jet power. In panel~(a), the jet-head velocity increases with jet power and, in the case with MF effects, gradually converges toward the nonmagnetized limit as the jet power rises. This reflects the weakening of MF effects as the jet power rises. When the power approaches $\sim 10^{48}\,\mathrm{erg/s}$, MF effects on the jet-head velocity become almost negligible. The gray dash-dotted line marks $\beta_{\rm h,b}=0.5$. Since all displayed solutions have $\beta_{\rm h,b}>0.03$, the $\beta_{\rm h,b}=0.03$ boundary lies outside the plotted range. The half-opening angle of the jet head in panel~(c) exhibits a similar trend. In contrast, the breakout location of the jet head shows little dependence on jet power in panel~(b). This behavior is primarily attributed to the high disk density, such that variations in the jet-head velocity are too weak to substantially alter the optical depth threshold for breakout. Hence, breakout takes place very close to the disk surface.

For the cocoon, the lateral expansion velocity initially increases with jet power but exhibits a cutoff beyond a certain value of jet power. This cutoff is caused by the limitation imposed by the Alfv\'en speed. Consequently, similar limitations appear in the evolution of the cocoon lateral radius and half-opening angle. Additionally, in the relatively low-power regime ($10^{41}\mbox{--}10^{44}\,\mathrm{erg/s}$), magnetic energy dissipation dominates the cocoon energy budget, leading to an enhancement of the cocoon energy, which may produce more luminous cocoon cooling emission after breakout. In contrast, in the relatively high-power regime, the cocoon is powered primarily by the jet. The increase in jet-head velocity reduces the fraction of jet energy transferred to the cocoon, and therefore the cocoon energy is lower.

As discussed in Section~\ref{sub:EF}, the direct magnetic terms at the jet head are dynamically subdominant. To identify the physical origin of the modest increase in jet-head velocity, we compare key breakout-location quantities between the magnetized and nonmagnetized cases as functions of jet power, adopting \(M_{\rm BH} = 10^{7}\,M_{\odot}\). These include the jet-density ratio, $\rho_{\rm j,MF}/\rho_{\rm j,NMF}$, the jet ram-pressure ratio, $\Pi_{\rm j,MF}/\Pi_{\rm j,NMF}$, the ratio of the half-opening angle of the jet head, $\theta_{\rm h,MF}/\theta_{\rm h,NMF}$, and the jet-head velocity ratio, $\beta_{\rm h,MF}/\beta_{\rm h,NMF}$. Here, $\Pi_{\rm j} = \rho_{\rm j} h_{\rm j} c^2 \Gamma_{\rm j}^2 \Gamma_{\rm h}^2(\beta_{\rm j}-\beta_{\rm h})^2$ is the jet ram-pressure term entering the head pressure balance. The subscripts MF and NMF refer to the magnetized and nonmagnetized cases, respectively. As shown in Figure~\ref{fig7}, the magnetized case gives a smaller jet-head half-opening angle, $\theta_{\rm h,MF}/\theta_{\rm h,NMF}<1$, especially at low jet power. This indicates enhanced jet collimation driven by cocoon confinement. For a fixed jet power, the smaller jet-head cross section increases the jet density, as shown by $\rho_{\rm j,MF}/\rho_{\rm j,NMF}>1$. The enhanced density then increases the jet ram-pressure term at the head, yielding $\Pi_{\rm j,MF}/\Pi_{\rm j,NMF}>1$. This leads to an increase in the jet-head velocity, with $\beta_{\rm h,MF}/\beta_{\rm h,NMF}\gtrsim1$. However, the increase in the head velocity is relatively modest, because \(\beta_{\rm h}\) depends nonlinearly on the head pressure balance and approaches an upper limit as the head becomes relativistic. These comparisons show that the increase in $\beta_{\rm h}$ is mainly an indirect, collimation-driven effect. Namely, the disk MF suppresses the lateral expansion of the cocoon, strengthens jet collimation, increases the jet density and ram pressure, and thereby increases the jet-head velocity.

In this work, we assume $z_{\rm x} = z_{\rm h}$. If $z_{\rm x}$ is considerably smaller than \(z_{\rm h}\), the power of magnetic energy dissipation decreases. This leads to a reduced contribution of dissipated magnetic energy to the cocoon energy, and thus the cocoon energy decreases significantly in the low jet-power regime. Correspondingly, the lateral expansion velocity, lateral radius, and half-opening angle of the cocoon at the breakout location all decline relative to the \(z_{\rm x}=z_{\rm h}\) case. Therefore, the reconnection-powered contributions to the cocoon energy and luminosity obtained for \(z_{\rm x}=z_{\rm h}\) should be regarded as optimistic upper bounds. However, the dynamical parameters of the jet head at the breakout location show only minor variations. This arises because the modulation of jet properties is dominated by magnetic-pressure effects acting on the cocoon. As our analysis centers primarily on jet propagation within MADs, this assumption leaves our key conclusions largely unaltered.

\subsection{Jet-Cocoon Breakout Emission}
We calculate thermal emission associated with the shock breakout of the jet-cocoon system. We adopt a supermassive BH with $M_{\rm BH}=10^{7}\,M_{\odot}$ and assume that the jet is launched at $R=10^{2}\,R_{\rm g}$. Figure~\ref{fig8} presents the luminosity and temperature of the jet head at shock breakout. The symbols $L_{\rm h}$ and $T_{\rm h}$ denote the breakout luminosity and temperature evaluated with the nonrelativistic, mildly relativistic, or relativistic prescriptions according to $\beta_{\rm h,b}$.

As shown in the left panel of Figure~\ref{fig8}, at low jet powers, the breakout luminosity with MF effects is higher than that without MF effects. However, this trend reverses at high jet powers, where the nonmagnetized breakout luminosity is typically $\sim 2$ times the magnetized value. The solid purple dot marks the point at which the nonmagnetized case reaches $\beta_{\rm h,b}=0.5$ and enters the fully relativistic regime, whereas the case with MF effects satisfies $\beta_{\rm h,b}>0.5$ throughout the jet-power range considered. Therefore, the pronounced separation between the two curves in the low-jet-power regime arises primarily because the jet heads fall into different breakout-velocity regimes and are consequently evaluated using different emission prescriptions; the resulting luminosity contrast should therefore be interpreted with caution, as it originates from the most model-dependent part of our calculation. At high jet powers, the reversal stems from a substantially smaller jet-head cross-sectional area $\Sigma_{\rm h,b}$ in the case with MF effects due to cocoon confinement, which in turn yields a lower initial breakout energy $E_0$. The corresponding temperatures converge in this regime, as illustrated in the right panel. It should be emphasized that these results depend on our one-dimensional analytic treatment, and multidimensional jet-cocoon interactions could modify them.

\begin{figure}[t]
\centering
\includegraphics[width=0.9\linewidth, height=0.7\linewidth]{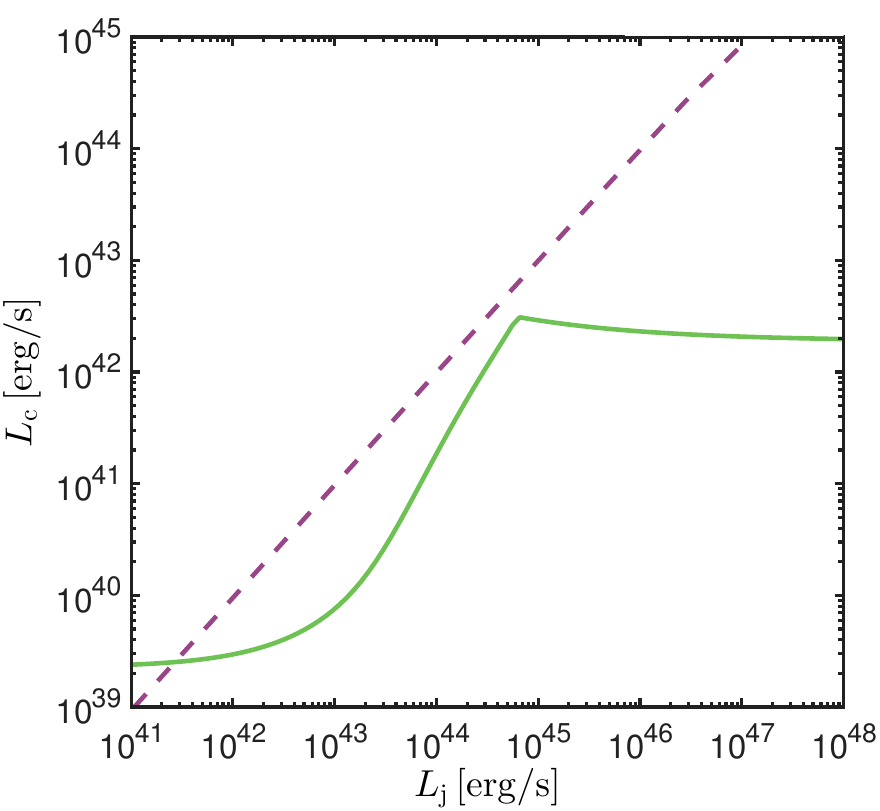}
\caption{Cocoon luminosity at the breakout location as a function of jet power in a MAD with $M_{\rm BH} = 10^{7}\,M_{\odot}$. Solid and dashed lines denote the results obtained with and without MF effects, respectively. The radial location of jet launching is fixed at $R = 10^{2}\,R_{\rm g}$.}
\label{fig9}
\end{figure}

Unlike the jet-head shock-breakout luminosity, the cocoon luminosity at the breakout location is effectively suppressed by the disk MFs, as shown in Figure~\ref{fig9}. This behavior indicates that the suppressive effect of magnetic pressure dominates over the contribution from magnetic energy dissipation for the cocoon luminosity. In addition, a luminosity plateau appears in the evolution curves of the cocoon luminosity at relatively high jet powers, which can be attributed to the limitation imposed by the Alfv\'en speed.

\begin{figure}[t]
\centering
\includegraphics[width=0.9\linewidth, height=0.7\linewidth]{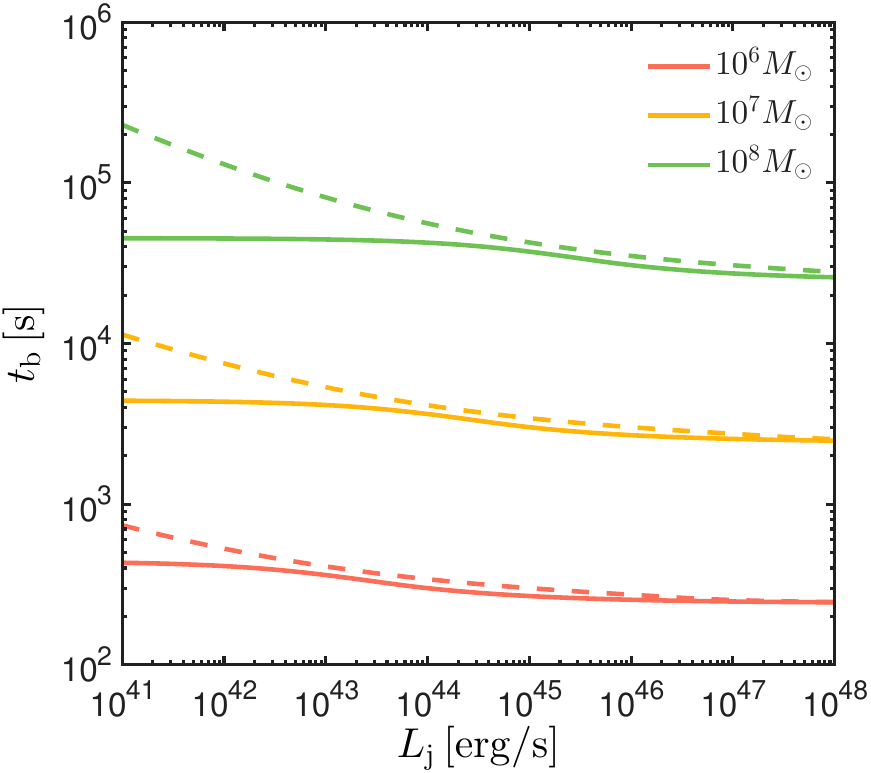}
\caption{Breakout time of jet-head shocks as a function of jet power. The red, yellow, and green lines denote the results obtained with $M_{\rm BH} = 10^{6}$, $10^{7}$, and $10^{8}\,M_{\odot}$, respectively. Solid and dashed lines denote the results obtained with and without MF effects, respectively. The radial location of jet launching is fixed at $R = 10^{2}\,R_{\rm g}$.}
\label{fig10}
\end{figure}

Figure~\ref{fig10} shows the breakout time of jet-head shocks as a function of jet power, defined as the interval from jet launching to shock breakout. First, irrespective of whether MF effects are included, the breakout time decreases as the jet power increases. In addition, MF effects can shorten the breakout time at relatively low jet powers, and this effect becomes more pronounced for higher supermassive BH masses. This implies a shorter delay between gravitational-wave and electromagnetic signals than in the nonmagnetized case.

\section{Conclusions and Discussion}
We have investigated the influence of disk MFs on jet propagation in MAD environments. In our parameterized model, the large-scale poloidal MFs mainly suppress the lateral expansion of the cocoon, leading to stronger jet collimation, a higher jet density, and an enhanced jet ram pressure at the head. As a result, the jet-head velocity can be modestly increased, particularly for relatively low jet powers. In this regime, the thermal luminosity from jet-head shock breakout is enhanced, though this enhancement is sensitive to the adopted regime-dependent emission prescriptions, while the breakout time is moderately shortened. These results suggest that large-scale poloidal magnetic fields in MADs may facilitate the breakout of low-power jets by confining the cocoon.

We emphasize that this conclusion should be interpreted cautiously in comparison with prior magnetized-jet simulations. \citet{Bromberg2016MNRAS} investigated intrinsically magnetized, Poynting-flux-dominated GRB jets, showing that global kink modes lower jet propagation speeds. Weakly magnetized simulations by \citet{Gottlieb2020MNRAS} demonstrate that even subdominant toroidal magnetic fields suppress instability-driven mixing between the jet and cocoon, weaken the jet-cocoon interface, and reshape post-breakout structures. Subsequent three-dimensional simulations tracked magnetically launched collapsar jets from the black hole out to breakout, revealing that magnetic dissipation and jet-stellar mixing substantially reduce jet magnetization, yielding weakly magnetized or hybrid-composition outflows at large radii \citep{Gottlieb2022MNRAS}. These works mainly focus on magnetic fields carried by the jet itself, whereas our analytic model assumes a nonmagnetized jet and isolates the effect of large-scale ordered MFs in MADs. Therefore, the modest increase in the jet-head velocity found here should not be interpreted as a generic magnetic acceleration of jets, but rather as an indirect consequence of cocoon confinement by MAD MFs.

Binary stellar-mass BH mergers may serve as a potential channel for launching low-power jets. Assuming that the merger remnant BH moves relative to the disk gas due to the recoil kick, its accretion process can be approximated by Bondi-Hoyle-Lyttleton (BHL) accretion \citep{Hoyle1939sc,Bondi1952mnras,Edgar2004}. For simplicity, we adopt $L_{\rm j}=\eta_{\rm j}\dot{M}_{\rm BHL}c^2$, motivated by Blandford--Znajek-like energetics \citep{Blandford1977MNRAS}, as a phenomenological scaling for the available jet power. The parameter \(\eta_{\rm j}\) is the efficiency for converting the accretion power into the jet kinetic luminosity. The BHL accretion rate is approximated as $\dot{M}_{\rm BHL}\simeq 4\pi G^2 M_{\rm r}^2\rho_{\rm d}/v_{\rm k}^3$, where \(M_{\rm r}\) and \(v_{\rm k}\) are the mass and kick velocity of the remnant black hole, respectively \citep{Edgar2004}. For the fiducial parameters $\eta_{\rm j}=0.1$, $M_{\rm r}=10^2\,M_{\odot}$, and $\rho_{\rm d}=10^{-11}\,\mathrm{g\,cm^{-3}}$, the corresponding jet power is $L_{\rm j} \sim 10^{43}\mbox{--}10^{45}\,\mathrm{erg/s}$ for $v_{\rm k} \sim 10^7\mbox{--}5\times10^7\,\mathrm{cm\,s^{-1}}$, under the assumption that the remnant remains embedded in the disk for longer than the engine duration. Such a range of jet power is well suited for exploring the influence of MAD MFs on jet-cocoon systems. Furthermore, if a stellar-mass binary BH merger occurs within an LLAGN disk and the central engine provides a sustained energy supply sufficient to power the jet up to breakout, the associated electromagnetic counterpart may appear as an X-ray flare. Meanwhile, a relatively short time delay is expected between the gravitational-wave signal and the associated electromagnetic signal.

In addition to MADs, several other astrophysical environments may also exhibit non-negligible MF effects during jet propagation. Supernova remnants are generally characterized by low average magnetization ($\sigma \ll 1$), suggesting that MFs are not globally dominant. However, localized regions such as shock-compressed layers and clump boundaries may reach moderate magnetization $\sigma \sim 10^{-4}\mbox{--}10^{-2}$ \citep[e.g.,][]{Inoue2012APJ}. When jets interact with these layers and clumps, MFs could influence jet propagation. Pulsar wind nebulae, powered by relativistic winds from pulsars, typically exhibit magnetization levels of $\sigma \sim 10^{-3}\mbox{--}10^{-1}$ \citep[e.g.,][]{Gaensler2006ARAA}. In such environments, the ambient MFs might affect the emission signatures of jet-driven forward shocks \citep{Granot2003ApJ}. Galactic nuclei are complex environments containing both dark matter and MFs. Jets propagating in such regions have their radiation luminosity modulated by interactions with dark matter \citep{Huang2020ApJ}. In addition, magnetized filamentary structures in galactic nuclei with $\sigma \sim 10^{-3}\mbox{--}10^{-1}$ (e.g., \citealt{Yusef-Zadeh1987ApJ}) may imply ordered large-scale MFs, which could influence the polarization signatures of jet emission \citep{Teboul2021MNRAS}. For binary systems hosting a magnetar companion, the ambient magnetization can be very high ($\sigma \gg 1$). If a jet is launched from the other binary component, the highly magnetized environment may substantially affect jet dynamics and emission. For instance, in GRBs, energy carried by magnetic winds can be injected into the jet, giving rise to a plateau feature in the afterglow light curve \citep{Dai1998AA,Li2026ApJ}.

In this work, we have not considered the scenario of choked jets \citep[e.g.,][]{Zhu2021ApJL} or calculated the associated nonthermal emission \citep[e.g.,][]{Tagawa2023ApJa}. In addition, jet dynamics are evaluated only up to the breakout location. However, both jet dynamics and emission should remain affected by disk MFs even after breakout. For instance, the large-scale ordered MFs outside MADs might influence the polarization signatures of the nonthermal emission. These aspects will be explored in future work.

\begin{acknowledgments}
We thank Hao-Qiang Zhang and Zhi-Lin Chen for helpful discussions. This work was supported by the National Natural Science Foundation of China (Grant Nos. 12494572, 12503052, 12473046, and 12221003) and the Guangxi Natural Science Foundation (Grant No. 2026GXNSFBA00640065).
\end{acknowledgments}

\end{document}